# Investigating the association between meteorological factors and the transmission and fatality of COVID-19 in the US

Meijian Yang

Department of Civil and Environmental Engineering, and Institute of the Environment, University of Connecticut, Storrs, Connecticut, USA

**Abstract**: A novel coronavirus disease (COVID-19) is sweeping the world and has taken away thousands of lives. As the current epicenter, the United States has the largest number of confirmed and death cases of COVID-19. Meteorological factors have been found associated with many respiratory diseases in the past studies. In order to understand that how and during which period of time do the meteorological factors have the strongest association with the transmission and fatality of COVID-19, we analyze the correlation between each meteorological factor during different time periods within the incubation window and the confirmation and fatality rate, and develop statistic models to quantify the effects at county level. Results show that meteorological variables except maximum wind speed during the day 13 – 0 before current day shows the most significant correlation ($P < 0.05$) with the daily confirmed rate, while temperature during the day 13 – 8 before are most significantly correlated ($P < 0.05$) with the daily fatality rate. Temperature is the only meteorological factor showing dramatic positive association nationally, particularly in the southeast US where the current outbreak most intensive. The influence of temperature is remarkable on the confirmed rate with an increase of over 5 pmp in many counties, but not as much on the fatality rate (mostly within ±0.01%). Findings in this study will help understanding the role of meteorological factors in the spreading of COVID-19 and provide insights for public and individual in fighting against this global epidemic.

1. Introduction

Since December 2019, a novel coronavirus disease (COVID-19) caused by severe acute respiratory syndrome coronavirus 2 (SARS-CoV-2) quickly spread over the world and is identified as a global epidemic by World Health Organization (WHO) (Chen et al., 2020; Gorbalenya et al., 2020; WHO, 2020a; F. Wu et al., 2020; Xu et al., 2020). Similar to other respiratory syndromes such as Severe Acute Respiratory Syndrome (SARS) and Middle East Respiratory Syndrome (MERS), the typical clinical symptoms of COVID-19 include fever, cough, fatigue, diarrhea, respiratory symptoms, kidney and liver failure, etc. (Bashir et al., 2020; Y. Wang et al., 2020). In general, COVID-19 has an incubation window of up to 14 days (Lauer et al., 2020; Tosepu et al., 2020), and the elderly patients with chronic comorbidities are more prone to be infected and killed by COVID-19 than healthy young people (Li et al., 2020; Qin et al., 2020; Yuki et al., 2020). Despite a much lower fatality rate compared with SARS and MERS, COVID-19 has killed far more people than the two combined (Kim and Goel, 2020; Mahase, 2020; D. Wang et al., 2020). Until June 28, 2020, this disease has infected nearly 10 million people and caused almost 500 thousand deaths in 213 countries (WHO, 2020b).

Meteorological factors played an important role in the spread of historical pandemics (Liu et al., 2020; Prata et al., 2020). Temperature, humidity and wind speed could affect the survival and transmission rates of virus such as SARS and influenza (Metz and Finn, 2015; Shaman et al., 2011; Tan et al., 2005; Yuan et al., 2006). Some studies indicate that temperature and diurnal temperature range are associated with the respiratory mortality (de Araujo Pinheiro et al., 2014; Luo et al., 2013). Due to the insufficient data and understanding, findings regarding the impact of weather factors on COVID-19 could be controversial. For example, different studies indicate that temperature could have positive, negative or no association with COVID-19 transmission(Briz-Redón and Serrano-Aroca, 2020; Ma et al., 2020; Shi et al., 2020; Xie and Zhu, 2020; Yao et al., 2020).

Soon after China and Europe, the United States becomes the epicenter of the COVID-19. So far, the US has 2.45 million confirmed cases and 125 thousand death cases, accounting for ~25% of global infected and death cases (WHO, 2020b). From late March on, the US federal and state governments start to take actions such as the stay-at-home and social distancing orders to slow down the reproduction and transmission of the virus (Engle et al., 2020; Lewnard and Lo, 2020). However, followed by the first outbreak centered in the New York metropolitan area, the current outbreak centered in multiple southern states appears to be even more serious (Bashir et al., 2020). Hence, understanding the influences of environmental factors on the transmission and fatality rate of COVID-19 is helpful for the optimization of medical resources and public and individual decision-makings to cope with the current and potential future outbreaks.

Some searches have tried to reveal and quantify such influences during the past months (Bashir et al., 2020; Liu et al., 2020; Ma et al., 2020; Prata et al., 2020; Tosepu et al., 2020; Y. Wu et al., 2020; Xie and Zhu, 2020). However, those studies either focus on one or several cities, or at a very coarse spatial resolution. Besides, as the current epicenter, studies regarding the impact of weather factors on COVID-19 in the US are relatively scarce. To fill those gaps, this study investigates the association between meteorological factors and the COVID-19 transmission and fatality rate in the US at county level, aiming at answering the following question: during which period of time do weather factors have the most significant correlation with the COVID-19 transmission and fatality? How does this association behave at different locations?

## 2. Data and Methods
### 2.1. Data sources

The data used in this study include weather data from the National Oceanic and Atmospheric Administration (NOAA) Local Climatological Data (LCD) (https://www.ncdc.noaa.gov/cdo-web/datatools/lcd), COVID-19 data from the Center for Systems Science and Engineering (CSSE) at John Hopkins University (JHU) (https://github.com/CSSEGISandData/COVID-19/), and the county population statistics in 2019 from The US Census Bureau (https://www.census.gov/data/datasets/time-series/demo/popest/2010s-counties-total.html). The NOAA LCD data provides hourly weather observations from nearly 2400 meteorological stations located in the US and the surrounding territories (Fig. 1). The weather factors consist of dry and wet bulb temperature, precipitation, relative humidity, visibility, and wind speed. The JHU COVID-19 data documents the daily cumulative confirmed and death cases in each county up to the current day. All data are publicly free available.

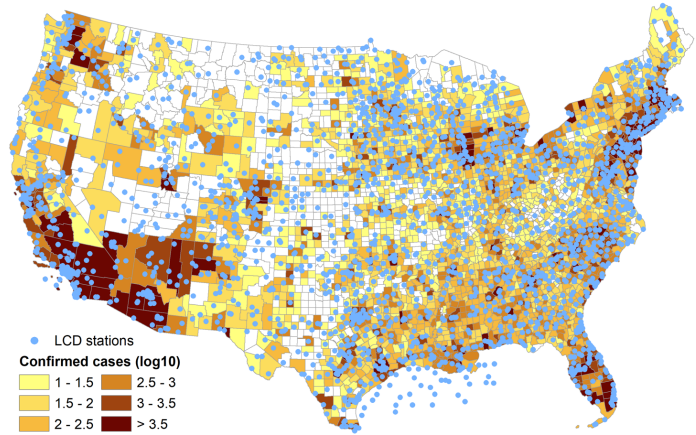

Figure 1. The location of LCD stations and the distribution of total confirmed COVID-19 cases until June 28, 2020

2.2. Methodology

Due to the limitations of current COVID-19 test ability, patients who have identical symptoms might be misdiagnosed (Yousefzai and Bhimaraj, 2020). Hence, we need to preprocess the daily reported data to make it consistent. We treat the data in June 28, 2020 as benchmark to correct the past confirmed and death numbers backwardly, making sure the numbers in the former days is no larger than in the later days. As for counties without LCD station located in, we take the nearest station to represent the weather condition. For counties with multiple LCD stations, we use the station that is closest to the county centroid.

Studies found that high population density could increase the risk of coronavirus transmission (Feng et al., 2020; Rocklöv and Sjödin, 2020). To account for this effect, we define the daily transmission rate (TR) as the ratio of the daily new confirmed cases (CC) to the county population (POP) (Equation 1). The definition of COVID-19 fatality rate varies with studies (Kim and Goel, 2020; Lipsitch, 2020; Spychalski et al., 2020). Considering the virus has a common incubation period of 14 days, we define the fatality rate (FR) as the ratio of the daily new death cases (DC) to the mean confirmed cases in the past 14 days (Equation 2). Studies show that a lagged effect of meteorological variables on reproduction and transmission of coronavirus could exist (Liu et al., 2020; Luo et al., 2013). We identify the period of time in the past 14 days when the correlation between meteorological variables and COVID-19 is strongest. The 14 meteorological factors are shown in Table 1. The meteorological variables during this period are taken for developing the statistic model. Due to the high variability among different counties, in terms of factors that may affect the transmission and fatality of COVID-19 such as hospital capacity, age structure, local policy, economic, custom, etc., for each county we fit a linear regression model to quantify the impact of weather factors (Equation 3).

$$TR_{i,t} = \frac{CC_{i,t}}{POP_i} \quad (1)$$

$$FR_{i,t} = \frac{DC_{i,t}}{\sum_{t-13}^{t} CC_{i,t}/14} \quad (2)$$

$$Y_i = \alpha_i + \beta_{i,l} X_{i,l} \quad (3)$$

Where $i$ represents the county, $t$ stands for the date, and $l$ is the number of meteorological variables.

Table 1. Daily meteorological factors

| ID | Description | Unit |
|---|---|---|
| 1 | Mean dry bulb temperature | °C |
| 2 | Maximum dry bulb temperature | °C |
| 3 | Minimum dry bulb temperature | °C |
| 4 | Total precipitation | Inch |
| 5 | Mean relative humidity | % |
| 6 | Mean visibility | % |
| 7 | Mean wet bulb temperature | °C |
| 8 | Maximum wet bulb temperature | °C |
| 9 | Minimum wet bulb temperature | °C |
| 10 | Mean wind speed | mph |
| 11 | Maximum wind speed | mph |
| 12 | Minimum wind speed | mph |
| 13 | Diurnal dry bulb temperature range | °C |
| 14 | Diurnal wet bulb temperature range | °C |

## 3. Results

3.1. Descriptive analysis

Figure 1 demonstrates the daily confirmed rate showing in the number of cases per million population (pmp) and the daily fatality rate (%). Counties with less than 10 total confirmed cases or 0 death cases until June 28 are excluded. The first record is in January 22, but the massive outbreak does not happen until March. The mean value of daily confirmed rate has been increasing in the past four months, from 0 to around 90 pmp, while the death rate is getting stable at around 5% after the surging in March. The 95% confidence interval doesn't show much departure from the mean, indicating the rate of COVID-19 transmission and fatality doesn't have notable variation from county to county.

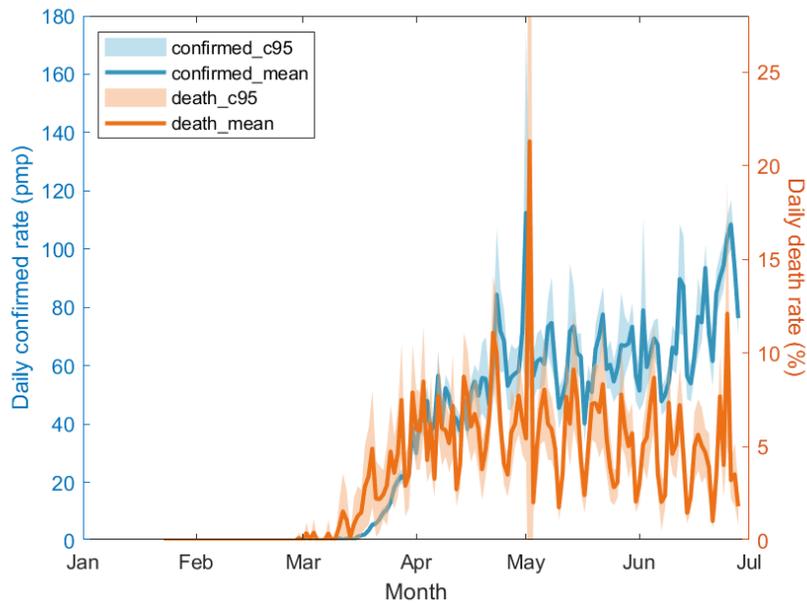

Figure 2. The mean and 95% confidence interval of daily COVID-19 confirmed and fatality rate across all counties.

3.2. Time period identification for the strongest correlation

In order to understand the time period during which COVID-19 are mostly associated with the meteorological factors, we performed a Pearson correlation analysis between each meteorological factor during different time period within the 14-day incubation window and the confirmed and fatality rate. Due to the very few reported cases in the first two months, data before March 1 are excluded in this analysis. Figure 3&4 indicate such correlation with confirmed and fatality rate at national scale. In terms of the daily confirmed cases (Figure 3), every meteorological factor has the highest correlation coefficient when taking the whole past two weeks (13 – 0 days before) into consideration, except for the maximum wind speed (7 – 0 days before). Regardless of the sign, all meteorological factors during this period are significantly correlated with the daily confirmed cases (P < 0.05), where temperature including both dry bulb and wet bulb has the most significant positive correlation (R > 0.7). Precipitation, visibility and diurnal dry bulb temperature range are positively correlated with the daily confirmed cases, while relative humidity, wind speed and diurnal wet bulb temperature range are negatively correlated with the daily confirmed cases.

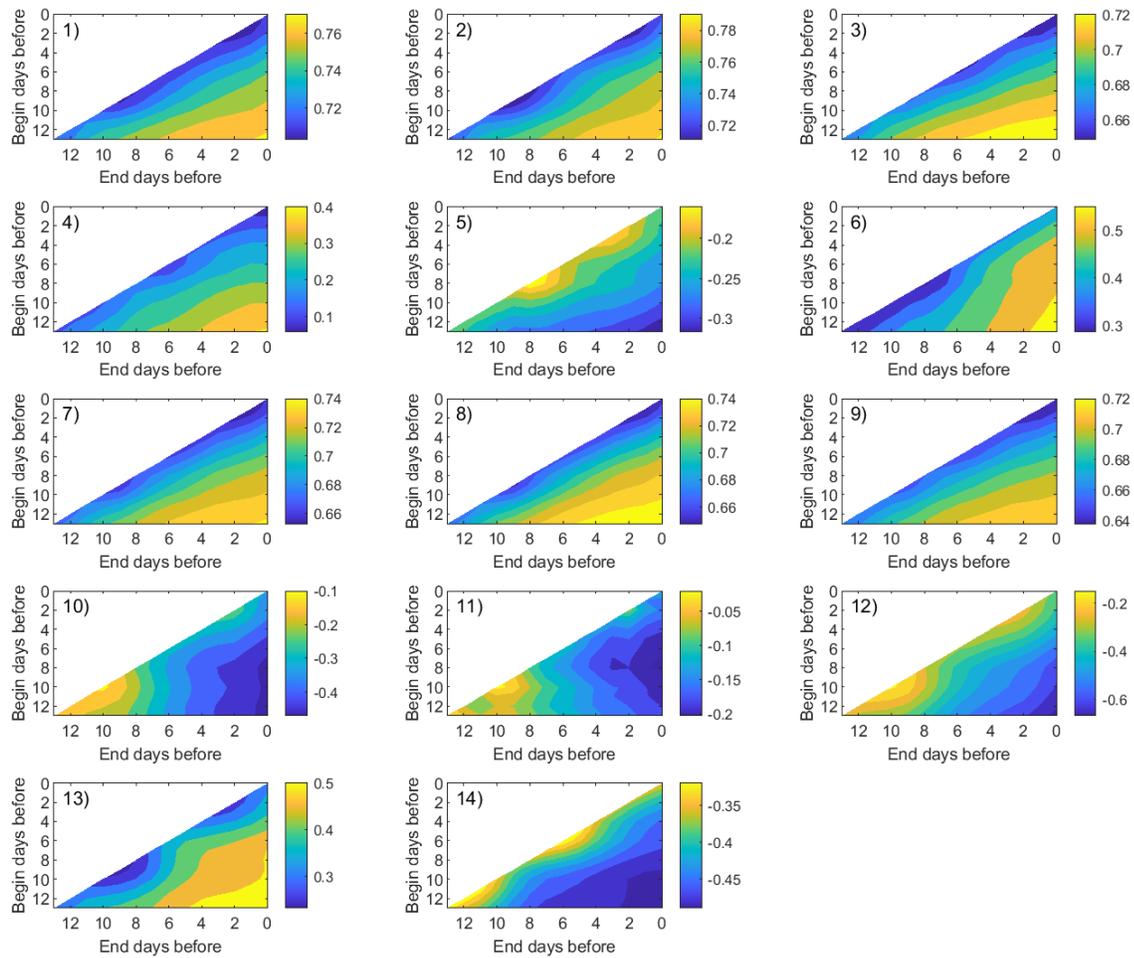

Figure 3. Correlation between meteorological factors at different lead time and daily confirmed rate (R = ±0.1794, P < 0.05). The lead time ranges from 13 days before to the current day (0 day before). Figure numbers are the weather factor IDs shown in Table 1.

Comparatively, the time period with the highest correlation between meteorological factors and fatality rate have some differences from confirmed rate. Figure 4 shows that precipitation, minimum wind speed and diurnal dry bulb temperature range during day 13 – 0 before are most significantly correlated with the fatality rate, while dry and wet bulb temperature from day 13 to day 8 before the current day has the highest correlation with the fatality rate. Other variables don't show significant correlation with the fatality rate. Besides, the correlation between meteorological factors and fatality rate is generally less significant than with confirmed rate. The correlation coefficient for temperature is around 0.28, while minimum wind speed has the highest correlation of -0.45 among all variables.

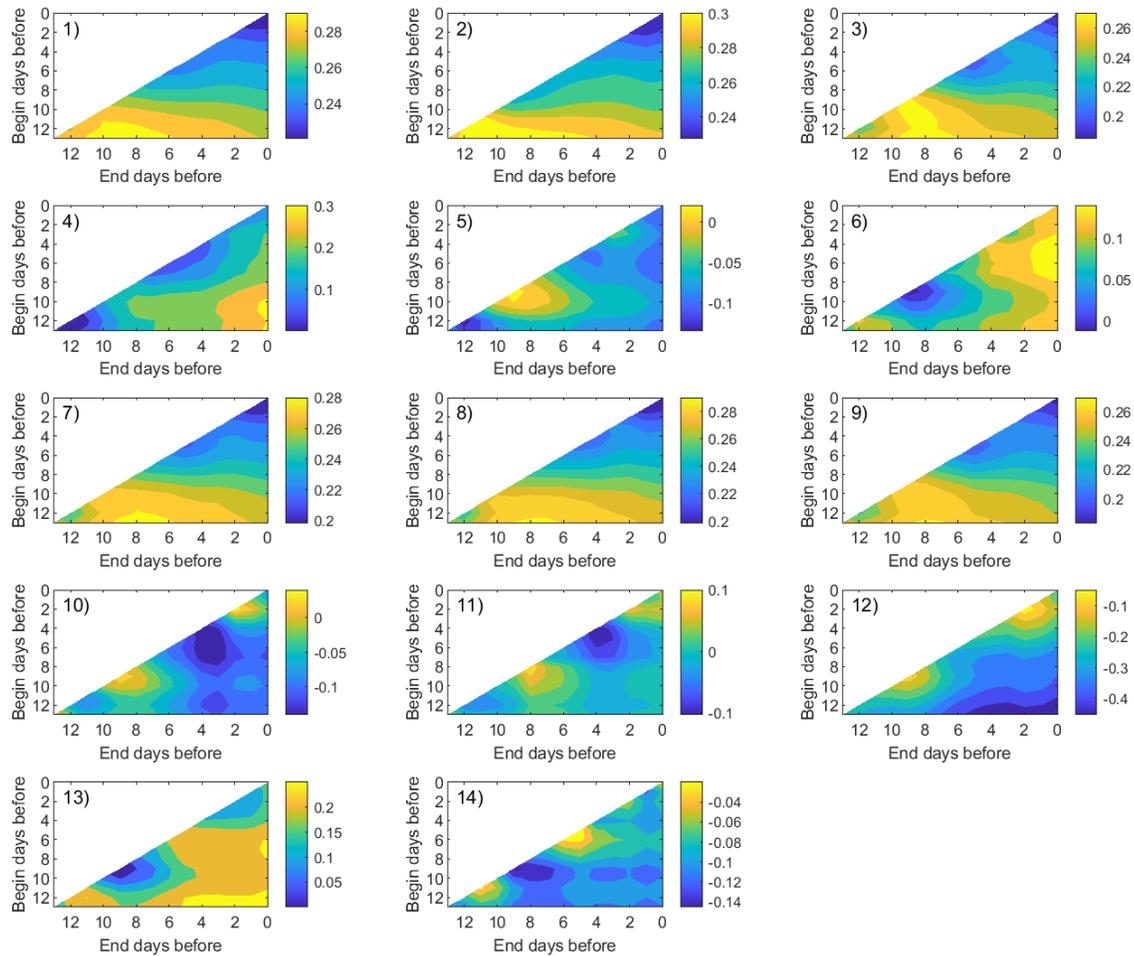

Figure 4. Correlation between meteorological factors at different lead time and daily fatality rate (R = ±0.1794, P < 0.05). The lead time ranges from 13 days before to the current day (0 day before). Figure numbers are the weather factor IDs shown in Table 1.

3.3. Quantitative analysis of the meteorological influence

To avoid the multicollinearity among independent variables in statistical models, we analyzed the correlation among different meteorological factors and removed the variables with strong multicollinearity (Figure S1). Besides, the time period with the highest correlation coefficient is selected for each meteorological factor and only the meteorological factors that are significantly correlated with COVID-19 (P < 0.05) are chosen as the independent variables. Hence, mean dry bulb temperature, total precipitation, mean relative humidity, mean visibility, mean wind speed, maximum wind speed, minimum wind speed and diurnal dry bulb temperature range are kept as the independent variables for the confirmed rate, while only dry bulb temperature, total precipitation, minimum wind speed and diurnal dry bulb temperature range are kept as the independent variables for the fatality rate.

Figure 5 shows the changes of confirmed and fatality rate per 1°C increase of the daily mean dry bulb temperature. 78% of the counties have an increase of confirmed rate, among which 33.4% of the county has an increasing rate of above 5 pmp and 5.5% of the county has an increasing rate of above 20 pmp (Figure 5a). Counties in the southeast US are most likely to have highest confirmed rate increase with temperature increase, which are also places where the current outbreak most intensive. 64.1% of the counties have an increase of fatality rate, but the changing magnitudes for most counties are less than ±0.01% (Figure 5b). Considering the actual confirmed and fatality rate (Figure 1), the influence of temperature on the confirmed rate is fairly remarkable while not as clear on the fatality rate. In contrast, the influences of precipitation, relative humidity and visibility on the confirmed rate are much smaller than temperature (Figure S2-S4). Wind speed and diurnal temperature range, however, show strong influence on the confirmed rate, but the influence has high spatial variability and the number of counties with increase or decrease rate are fairly close (Figure S5-S8). Similarly, the number of counties with increase or decrease fatality rate under one unit increase of precipitation, wind speed and diurnal temperature range are also very close (Figure S9-S11). Hence, nationally speaking, temperature is the only meteorological factor that has significant positive impact on the transmission and fatality of COVID-19.

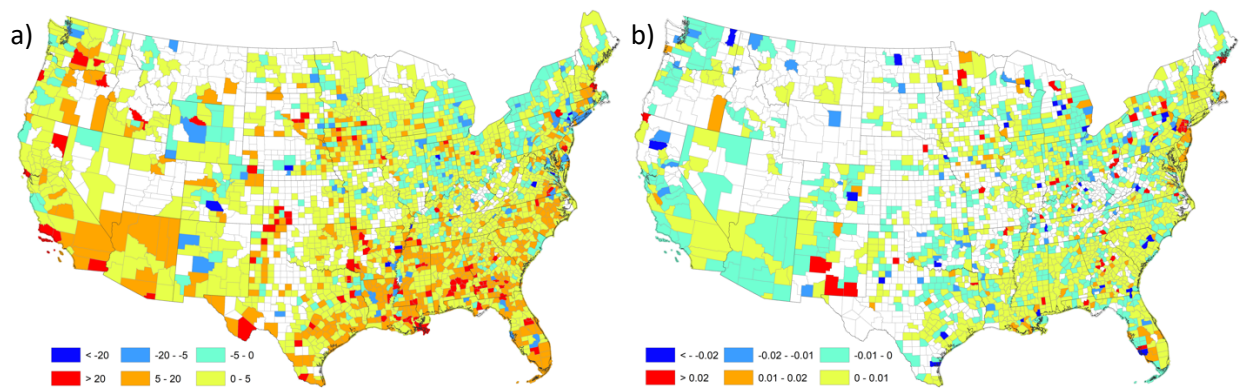

Figure 5. Changes of a) confirmed rate (pmp) and b) fatality rate (%) of COVID-19 in each county by 1 °C increase of the daily mean dry bulb temperature.

## 4. Discussions

Our results indicate that temperature has positive association with the daily confirmed rate, which is consistent with some past studies (Auler et al., 2020; Xie and Zhu, 2020) while disagrees with some other studies (Liu et al., 2020; Qi et al., 2020; Shi et al., 2020). Most of those studies were performed in China where most cases are reported during the winter time (January – March). In this regard, our study provides a more solid evidence as our data series has a much larger temperature span from winter to summer. Physiologically, it is possible that people in warm weather will be more likely to resist wearing protective equipment such as face masks, which could increase the risk of virus transmission. However, more dynamic investigations and experiments need to be done before we understand whether such association is biomedically solid or just a coincidence. On the other hand, the fatality rate is not as significantly associated with meteorological factors as the confirmed rate. This is primarily because of the continuous improvement of the test and treatment measures maintain the fatality rate at a relatively stable level when the confirmed rate is still climbing.

Our study has several limitations. First, we do not have a long time series. The length of record from the start of intensive outbreak in the US to the day that this study is performed is only 120 days (March 1 – June 28). As the epidemic continues, future evidence might show a different pattern. Second, it's hard to draw a conclusion that meteorological factors promote or inhibit the transmission and fatality rate of

COVID-19 only from the correlation and regression analysis. Hence, we only aim at presenting the association between these two, while further research is needed to address the innate cause and effect.

## 5. Conclusions

Weather factors plays a role in the fight against COVID-19 in the US and globally. This study investigates the association between meteorological factors and the transmission and fatality rate of COVID-19 at county level in the US. In general, meteorological variables except maximum wind speed during the day 13 – 0 before current day shows the most significant correlation with the daily confirmed rate, while temperature during the day 13 – 8 before are most significantly correlated with the daily fatality rate. In terms of the correlation with meteorological variables, the confirmed rate is much more affected than the fatality rate. The quantitative analysis demonstrates that temperature is the only meteorological variable showing an obvious pattern of significantly positively affecting the transmission and fatality of COVID-19. The influence of temperature is remarkable on the confirmed rate with an increase of over 5 pmp in many counties in the southeast US, but not as much on the fatality rate (mostly within ±0.01%). Other factors either have negligible impact or have similar number of counties that are positively or negatively affected. This study provides an insight of how and when meteorological factors are associated with COVID-19, and is helpful for the further understanding of the mechanism of such impacts.

**Declaration of competing interest**

The author declare that they have no known competing financial interests or personal relationships that could have appeared to influence the work reported in this paper.

**Acknowledgements**

**Appendix**

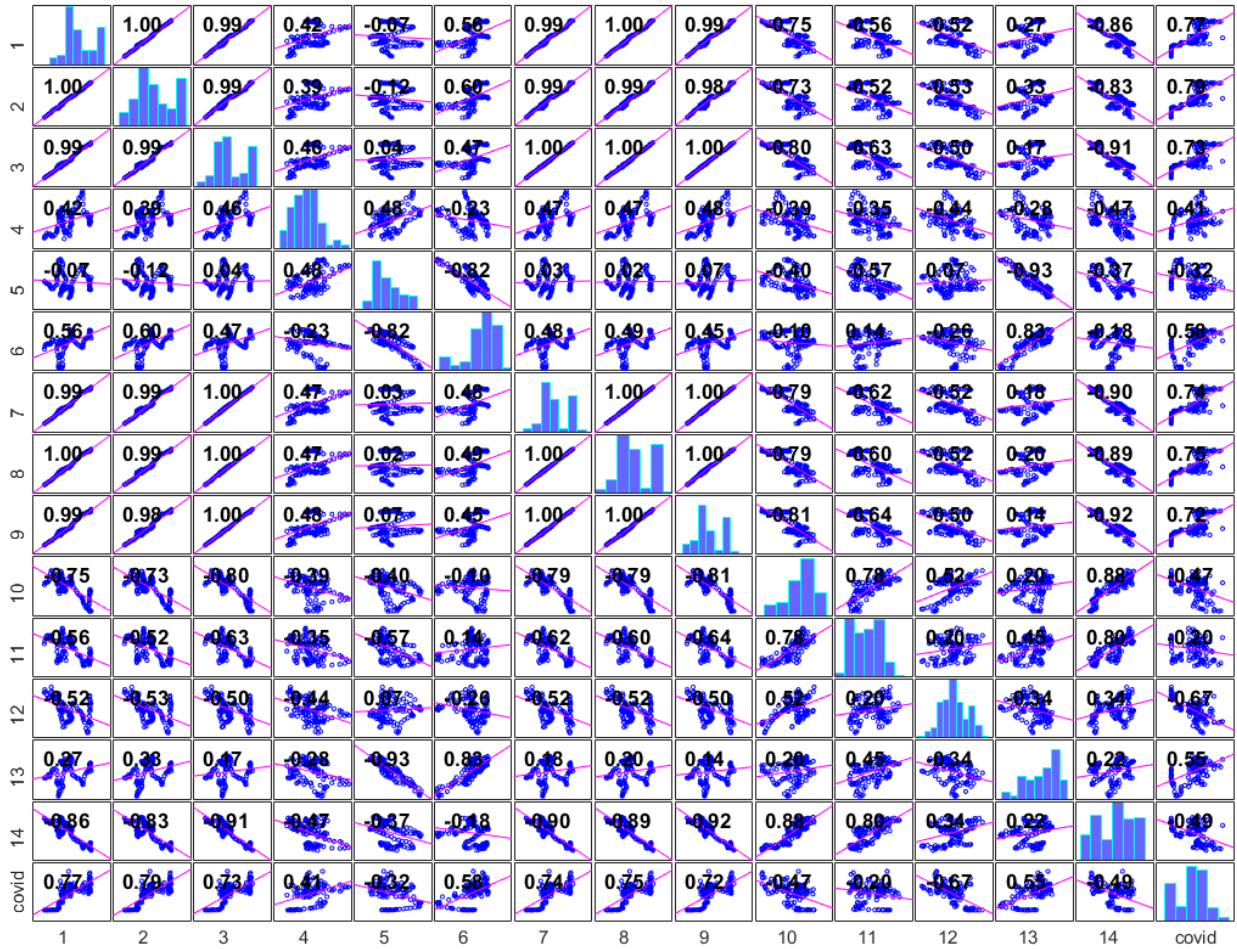

Figure S1. Correlation matrix among meteorological variables and COVID-19.

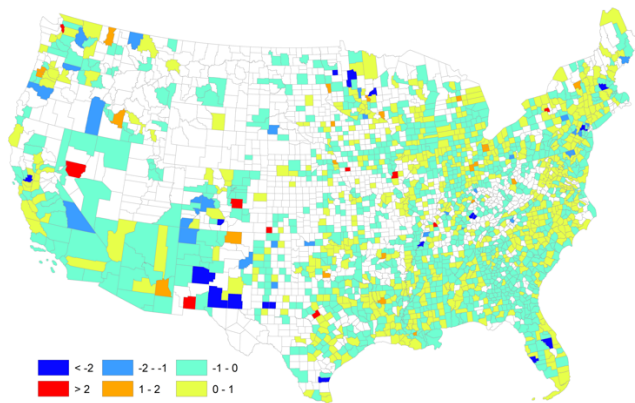

Figure S2. Changes of confirmed rate (pmp) of COVID-19 in each county by 1 inch increase of the total precipitation

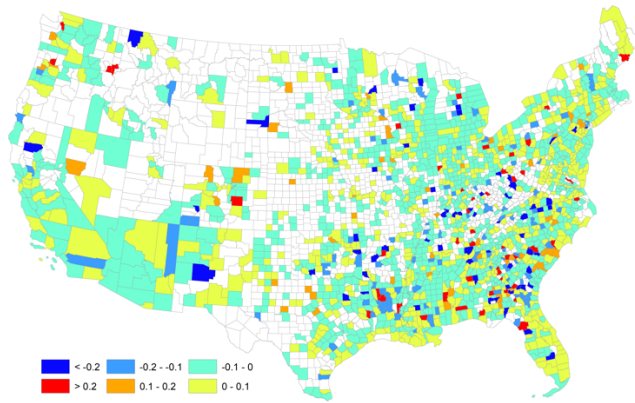

Figure S3. Changes of confirmed rate (pmp) of COVID-19 in each county by 1% increase of the mean relative humidity

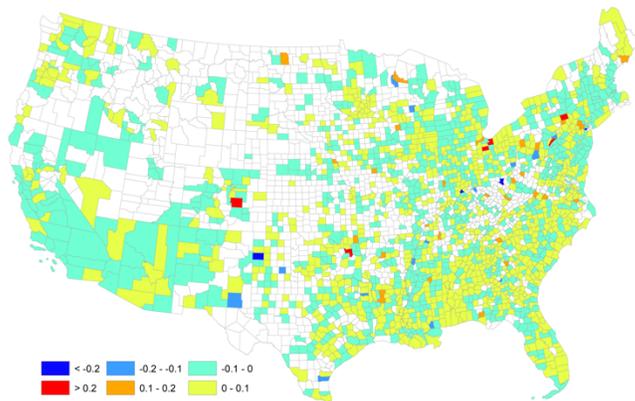

Figure S4. Changes of confirmed rate (pmp) of COVID-19 in each county by 1% increase of the mean visibility

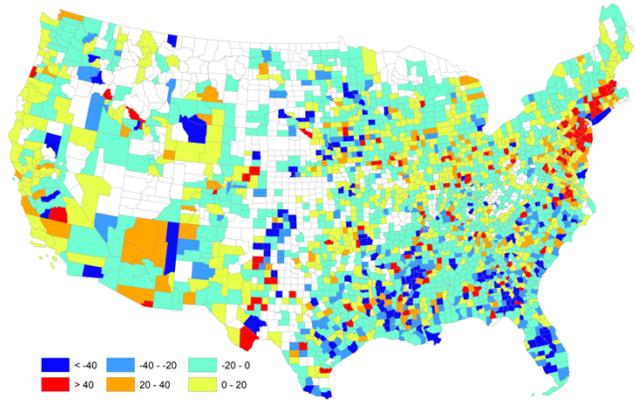

Figure S5. Changes of confirmed rate (pmp) of COVID-19 in each county by 1 mph increase of the mean wind speed

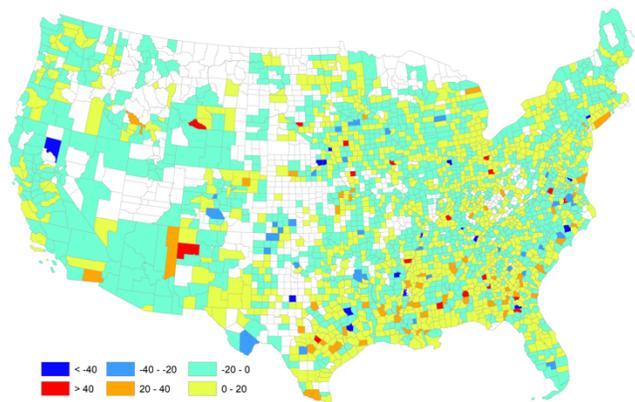

Figure S6. Changes of confirmed rate (pmp) of COVID-19 in each county by 1 mph increase of the maximum wind speed

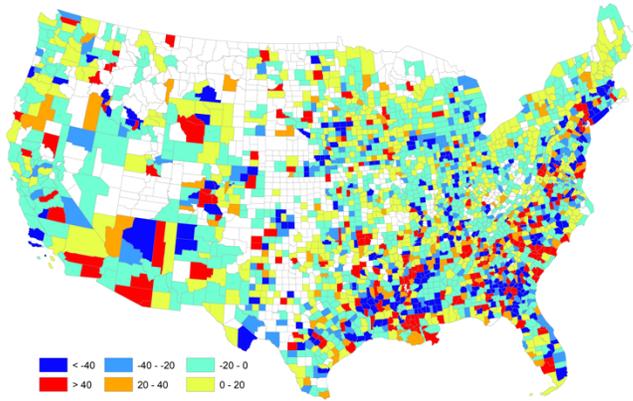

Figure S7. Changes of confirmed rate (pmp) of COVID-19 in each county by 1 mph increase of the minimum wind speed

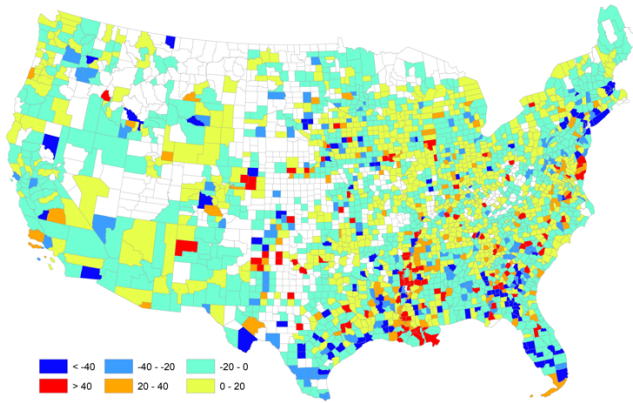

Figure S8. Changes of confirmed rate (pmp) of COVID-19 in each county by 1 °C increase of the diurnal dry bulb temperature range

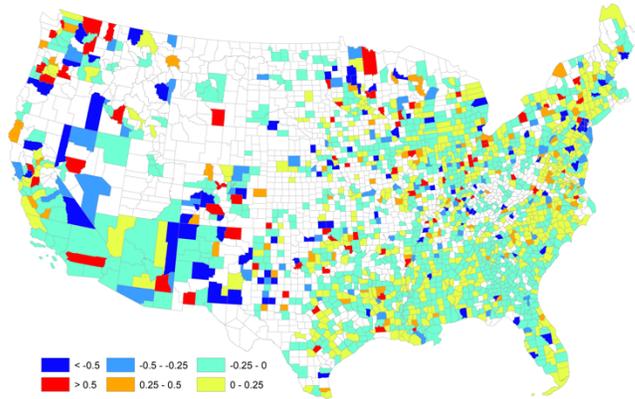

Figure S9. Changes of fatality rate (%) of COVID-19 in each county by 1 inch increase of the total precipitation

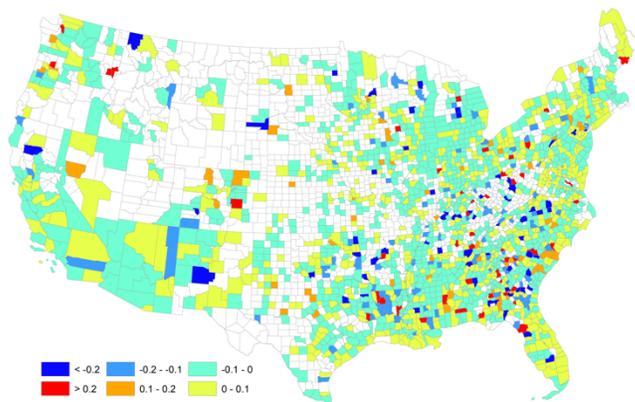

Figure S10. Changes of fatality rate (%) of COVID-19 in each county by 1 mph increase of the minimum wind speed

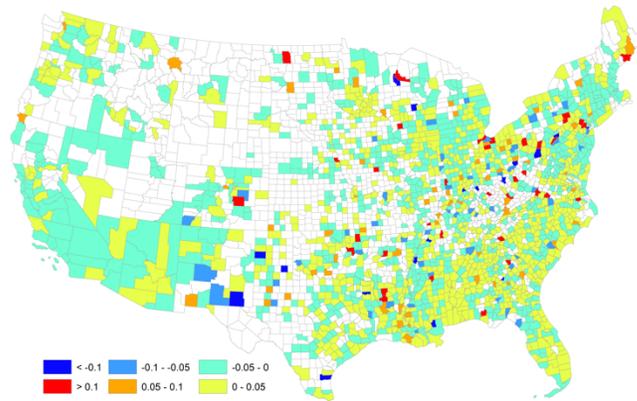

Figure S11. Changes of fatality rate (%) of COVID-19 in each county by 1 °C increase of the diurnal dry bulb temperature range